\documentclass[conference]{IEEEtran}
\IEEEoverridecommandlockouts
\usepackage{cite}
\usepackage{amsmath,amssymb,amsfonts}
\usepackage{algorithmic}
\usepackage{graphicx}
\usepackage{float}
\usepackage{subfigure}
\usepackage{textcomp}
\usepackage{xcolor}
\usepackage{multirow}

\def\BibTeX{{\rm B\kern-.05em{\sc i\kern-.025em b}\kern-.08em
    T\kern-.1667em\lower.7ex\hbox{E}\kern-.125emX}}
\begin{document}

\bstctlcite{IEEEexample:BSTcontrol}
\title{VESTA: A Versatile SNN-Based Transformer Accelerator with Unified PEs for Multiple Computational Layers\\
}

\author{\IEEEauthorblockN{Ching-Yao Chen, Meng-Chieh Chen, and Tian-Sheuan Chang}

\IEEEauthorblockA{\textit{Department of Electronics and Electrical Engineering, National Yang Ming Chiao Tung University,}
Hsinchu, Taiwan \\
Email: jameschen.c@nycu.edu.tw, chieh28.ee10@nycu.edu.tw, tschang@nycu.edu.tw
}
}

\maketitle

\begin{abstract}%

Spiking Neural Networks (SNNs) and transformers represent two powerful paradigms in neural computation, known for their low power consumption and ability to capture feature dependencies, respectively. However, transformer architectures typically involve multiple types of computational layers, including linear layers for MLP modules and classification heads, convolution layers for tokenizers, and dot product computations for self-attention mechanisms. These diverse operations pose significant challenges for hardware accelerator design, and to our knowledge, there is not yet a hardware solution that leverages spike-form data from SNNs for transformer architectures. In this paper, we introduce VESTA, a novel hardware design that synergizes these technologies, presenting unified Processing Elements (PEs) capable of efficiently performing all three types of computations crucial to transformer structures. VESTA uniquely benefits from the spike-form outputs of the Spike Neuron Layers \cite{zhou2024spikformer}, simplifying multiplication operations by reducing them from handling two 8-bit integers to handling one 8-bit integer and a binary spike. This reduction enables the use of multiplexers in the PE module, significantly enhancing computational efficiency while maintaining the low-power advantage of SNNs. Experimental results show that the core area of VESTA is \(0.844 mm^2\). It operates at 500MHz and is capable of real-time image classification at 30 fps.

\end{abstract}

\section{Introduction}

\begin{figure}[t]
\centering
\includegraphics[width=1.0\linewidth]{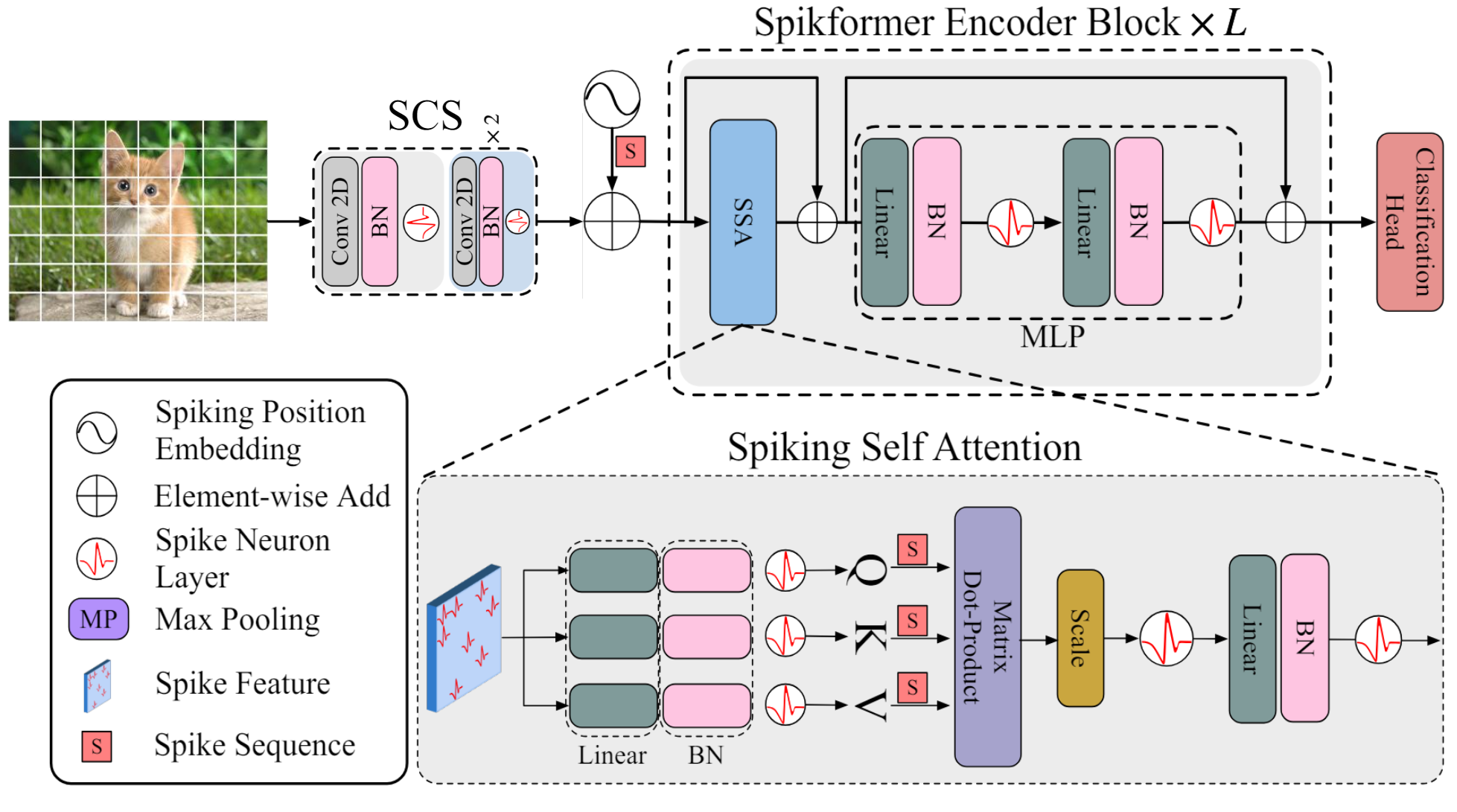}
\caption{The architecture of Spikformer V2 \cite{zhou2024spikformer} includes a Spiking Convolutional Stem module (SCS), a Spikformer encoder, and a Linear classification head.}
\label{spikformer}
\end{figure}

Transformer-based models have significantly advanced numerous applications due to their ability to capture long-range dependencies within data. However, deploying these complex models on edge devices remains challenging due to their high computational demands and the necessity for multiple computing elements for different types of computational layers. For instance, the "Spikformer V2" \cite{zhou2024spikformer} model, depicted in Fig.~\ref{spikformer}, aims to combine the energy efficiency of SNNs with the performance benefits of the self-attention mechanism. The Spiking Convolutional Stem (SCS) module in Spikformer V2 requires convolutional computation, while the Spikformer Encoder Blocks require dot-product and linear computations. Consequently, deploying transformer-based models on edge devices presents a significant design challenge.

This paper introduces VESTA, an efficient solution leveraging the "Spikformer V2-8-512-IAND" model, a variant of Spikformer optimized for edge computing by utilizing pure binary activation for inter-layer information propagation. VESTA addresses the computational needs of transformer architectures with several specialized operations. For convolution layers with spike-form inputs, VESTA employs Zig-Zag Spiking Convolution (ZSC), optimizing input placement to maximize the utilization of the processing element (PE) module. For linear layers, VESTA utilizes Weight Stationary Spiking Linear Operation (WSSL), efficiently managing computations by iterating through input maps across multiple timesteps, thus reducing memory overhead. In self-attention mechanisms, VESTA incorporates Spiking Tile-wise Dot Product Calculation (STDP), performing dot product computations immediately after completing columns of matrix \( V \), thereby reducing memory usage. For convolution layers with 8-bit inputs, VESTA employs Shift-and-Sum Spiking Convolution (SSSC), treating 8-bit inputs as eight 1-bit inputs and summing the results after appropriate shifts to perform 8-bit multiplications effectively.

A further innovation within the VESTA architecture is the integration of the Temporal Fused Leaky Integrate-and-Fire (TFLIF) module. This module efficiently handles spike-form outputs across multiple timesteps and integrates normalization processes directly into its computation. By applying the VESTA architecture to the ImageNet dataset across four timesteps, we effectively mitigate the accuracy loss typically associated with quantizing from float32 to uint8. Through these enhancements, VESTA aims to establish a new benchmark for deploying advanced neural network architectures in power-efficient environments.

\section{The VESTA Architecture and Operations} 
\begin{figure}[t]
\centering
\includegraphics[width=1.0\linewidth]{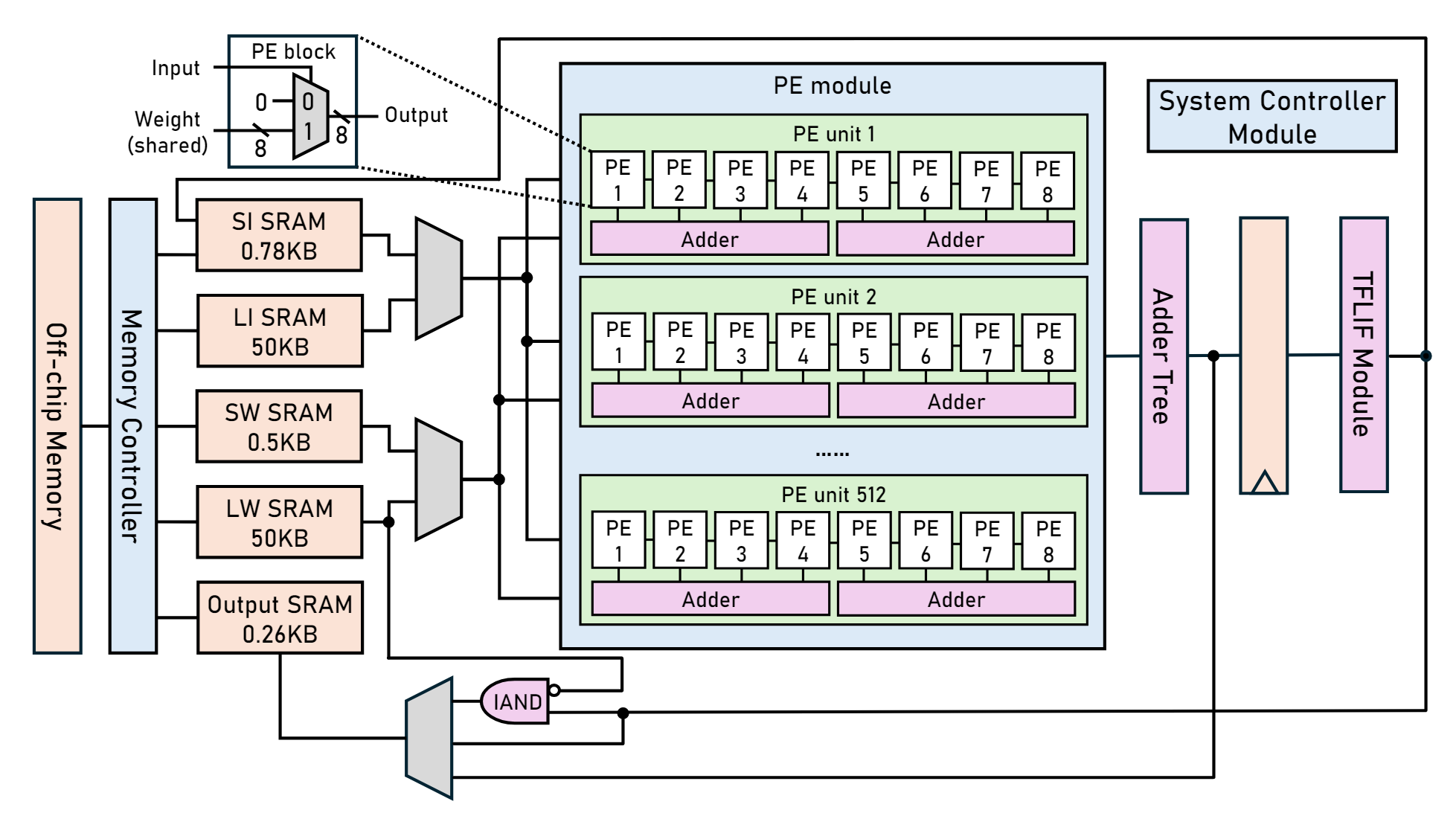}
\caption{VESTA overall architecture}
\end{figure}
This section examines the VESTA architecture and its capability to efficiently handle convolutional, linear, and dot product computations for transformer-based models, emphasizing its role in reducing complexity and enhancing energy efficiency.

\subsection{Architecture Overview}

At the heart of this architecture is the PE module, consisting of 512 units. Each unit handles one shared 8-bit weight and eight distinct 1-bit inputs, with each of the eight PEs within the unit utilizing the same weight but a unique input. By exploiting weight sharing, this configuration has proven to be efficient for parallel computations across multiple timesteps. 

The System Controller Module manages memory and the operation mode of the PE module, ensuring flexible and efficient handling of varying computational demands. 

The Large Weight (LW) SRAM and Small Weight (SW) SRAM store the 8-bit weights, tailored to different size requirements of the operations, while the Large Input (LI) SRAM and Small Input (SI) SRAM house the 1-bit inputs, facilitating efficient data access and processing.

An adder tree aggregates the outputs from the PEs and forwards them to the TFLIF module, where they are converted into single-bit (spike-form) outputs. These outputs are then stored in the Output SRAM, utilizing minimal space to enhance storage efficiency. This design significantly reduces delays associated with frequent access to off-chip memory, ensuring continuous operation and improved system performance. 

\subsection{The Temporal Fused LIF Module (TFLIF)}

\begin{figure}[t]
\centering
\includegraphics[width=1.0\linewidth]{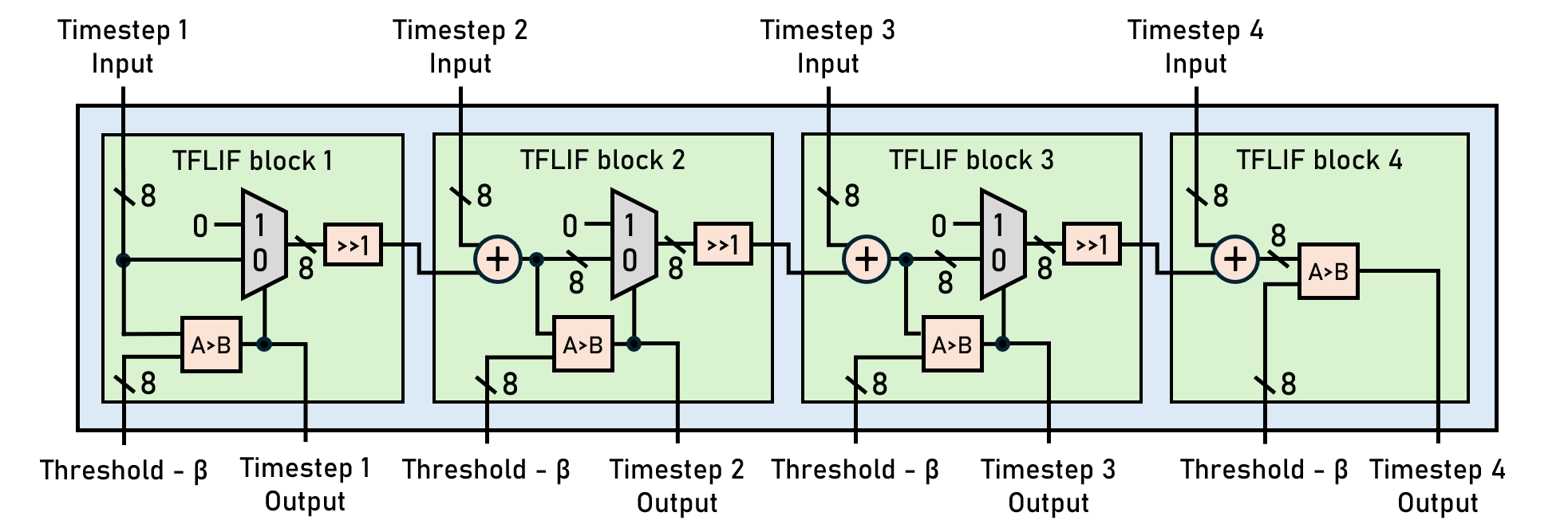}
\caption{The TFLIF module takes advantage of the integration of the bias value from the BN layer with the threshold value of the LIF layer, simplifying the computation process.}
\label{LIF}
\end{figure}

Spiking Neural Networks (SNNs) typically require multiple timesteps to achieve accuracy comparable to traditional Artificial Neural Networks (ANNs). In the Spikformer model we are implementing, a timestep of four is utilized. While employing multiple timesteps inherently increases computational demands, it concurrently offers the advantage of weight sharing, as the weight map remains constant across different timesteps. This characteristic facilitates the simultaneous calculation of outputs across all timesteps, thereby maximizing the utilization of the PE module.

As illustrated in Fig.~\ref{LIF}, the TFLIF module accepts four 8-bit outputs from the PE module or Adder tree and converts them into spike-form outputs. This conversion substantially reduces memory usage compared to processing each timestep individually. Additionally, in the Spikformer architecture, the LIF layer is consistently positioned after a batch normalization (BN) layer. We further optimize this arrangement by subtracting the threshold value of the LIF layer from the bias value in the BN layer, effectively integrating the normalization process directly into the LIF computation. This adjustment eliminates the need to calculate the BN layer separately, further enhancing the efficiency of the model.

\subsection{Zig-Zag Spiking Convolution (ZSC)}
\begin{figure}[t]
\centering
\includegraphics[width=0.9\linewidth]{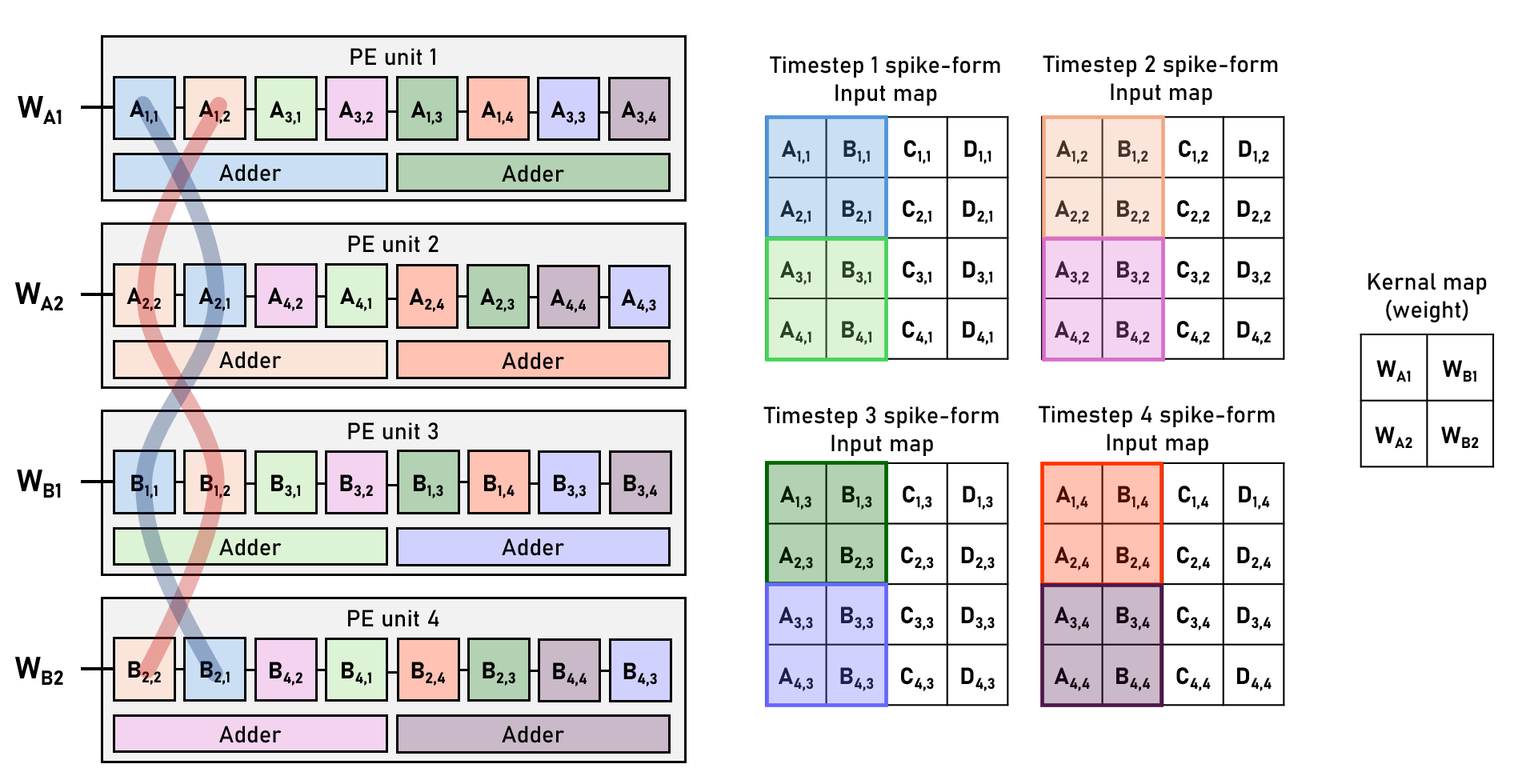}
\caption{The Zig-Zag Spiking Convolution (ZSC) positions each 2x2 input submatrix from consecutive timesteps in a unique zig-zag pattern to optimize the usage of processing elements (PE) within the module. This setup is shown in Fig.\ref{1-bit_conv}, where groups of two input submatrices are aligned to enable the simultaneous processing of two pixels from one channel over the four timesteps by every four PE units.}
\label{1-bit_conv}
\end{figure}

The Spiking Convolutional Stem (SCS) module utilizes 4 convolution layers with a 2x2 kernel map and a stride of 2. Leveraging the uniformity of weights across different timesteps and the inherent weight-sharing property of convolution layers, we have optimized the utilization of the PE module for convolution layers with spike-form inputs through the Zig-Zag Spiking Convolution (ZSC) technique. Since all weights remain consistent across the four timesteps, we strategically arrange groups of two 2x2 input submatrices and feed them into the PE module as depicted in Fig.~\ref{1-bit_conv}. This ZSC configuration enables the processing of outputs for two pixels in one channel over four timesteps for every four PE units. Subsequently, the outputs from each PE unit are summed across all input channels before being sent to the TFLIF module. This summation is essential, as it combines the individual contributions of each channel to form a complete output for each pixel. By eliminating the need to store intermediate outputs from the PE module, ZSC significantly reduces memory size and data transfer requirements. This strategy is particularly effective for layers with varying numbers of channels and different input map dimensions within the convolution layers, ensuring efficient data handling and processing across multiple layers within the SCS module.

\subsection{Shift-and-Sum Spiking Convolution (SSSC)}

\begin{figure}[t]
\centering
\includegraphics[width=0.9\linewidth]{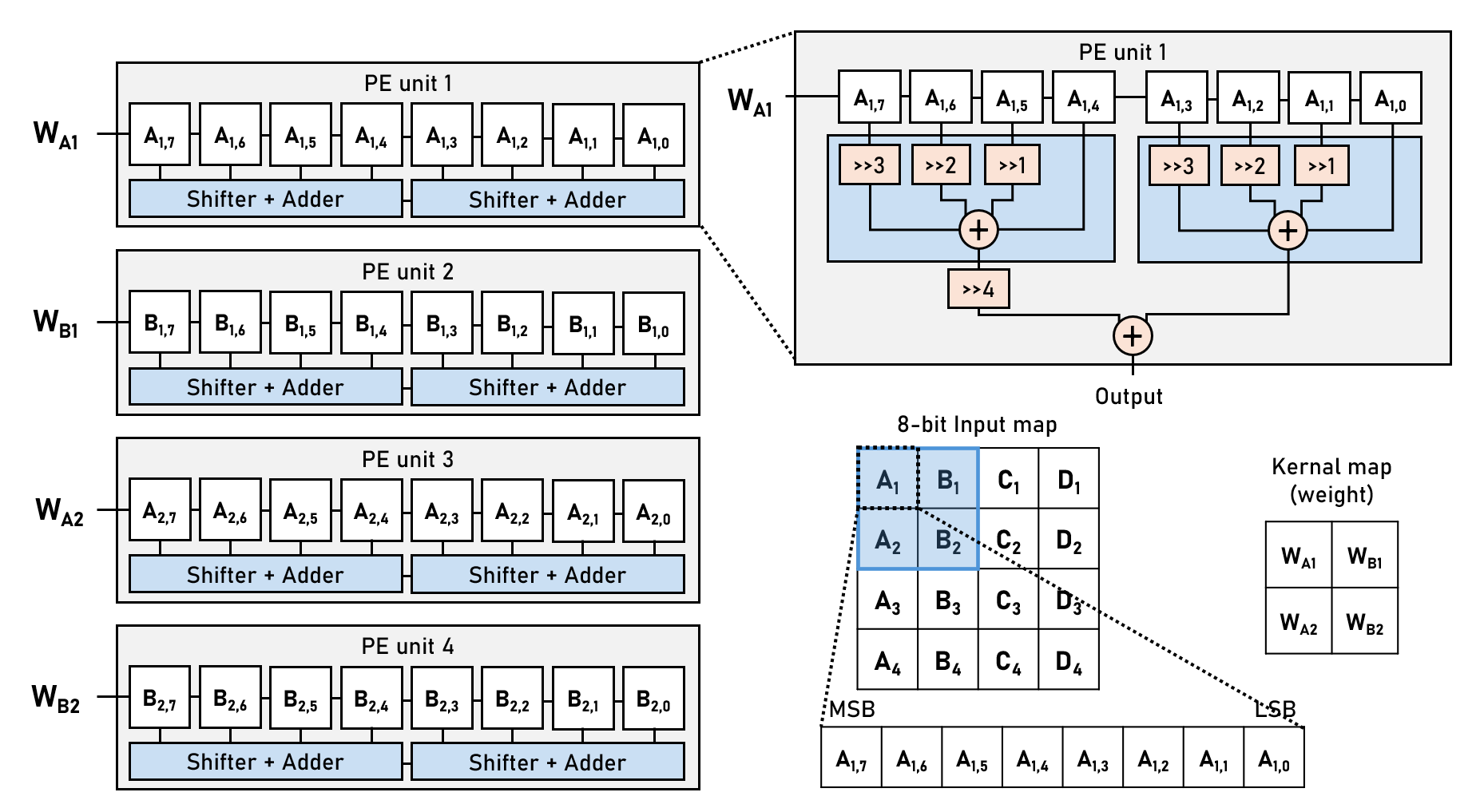}
\caption{The Shift-and-Sum Spiking Convolution (SSSC) method treats an 8-bit input as 8 spike-form inputs. The result of each PE block is shifted accordingly before summation.}
\label{8-bit_conv}
\end{figure}

While the majority of operations in "Spikformer V2" leverage spike-form inputs characteristic of each layer, the initial input layer presents a unique challenge due to its source of data. This layer, part of the Spiking Convolutional Stem (SCS) module, processes 8-bit images from the ImageNet dataset. Each pixel, represented as an 8-bit integer, contributes to a 2D convolution operation that manages a 224×224 image with 3 channels, using a kernel size of 2 and a stride of 2 for downsampling. The 8-bit nature of both the convolution layer's weights and its inputs necessitates specific modifications to the PE module to accommodate these higher-resolution operations.

The proposed Shift-and-Sum Spiking Convolution (SSSC) method addresses this challenge by adapting the weight-sharing technique and treating 8-bit inputs as eight 1-bit inputs. As depicted in Fig.~\ref{8-bit_conv}, this approach allows for the summation of results from each PE in the units after being shifted according to their positions in the grid, effectively producing an output equivalent to multiplying 8-bit inputs with 8-bit weights. Unlike ZSC, the SSSC method does not need to handle inputs from multiple timesteps since the input is the same image across all four timesteps.

\subsection{Weight Stationary Spiking Linear Operation (WSSL)}

\begin{figure}[t]
\centering
\includegraphics[width=0.8\linewidth]{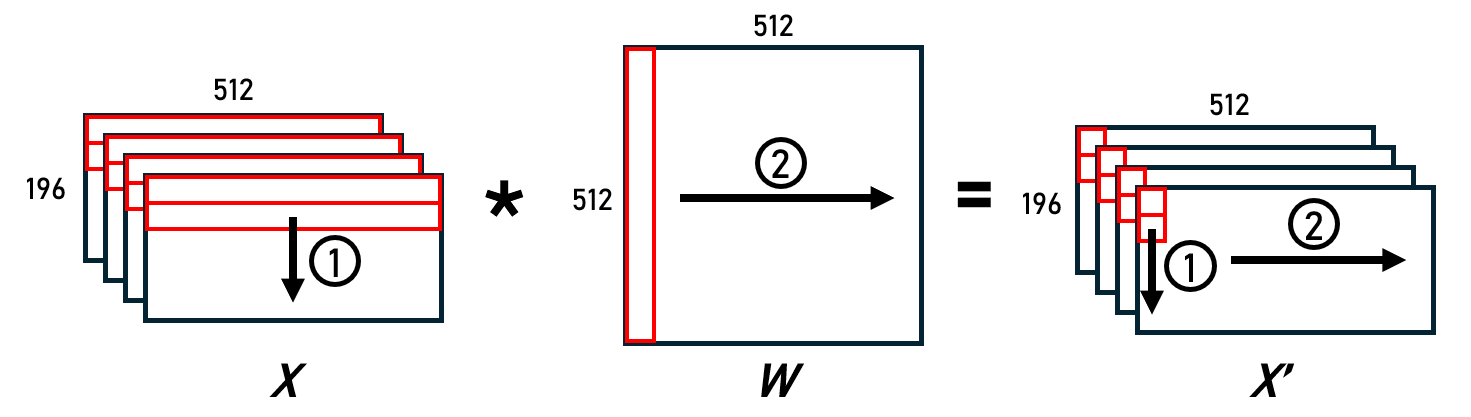}
\caption{Weight Stationary Spiking Linear Operation (WSSL) for linear layers with a height of 512.}
\label{linear_layer}
\end{figure}

\begin{figure}[t]
\centering
\includegraphics[width=0.8\linewidth]{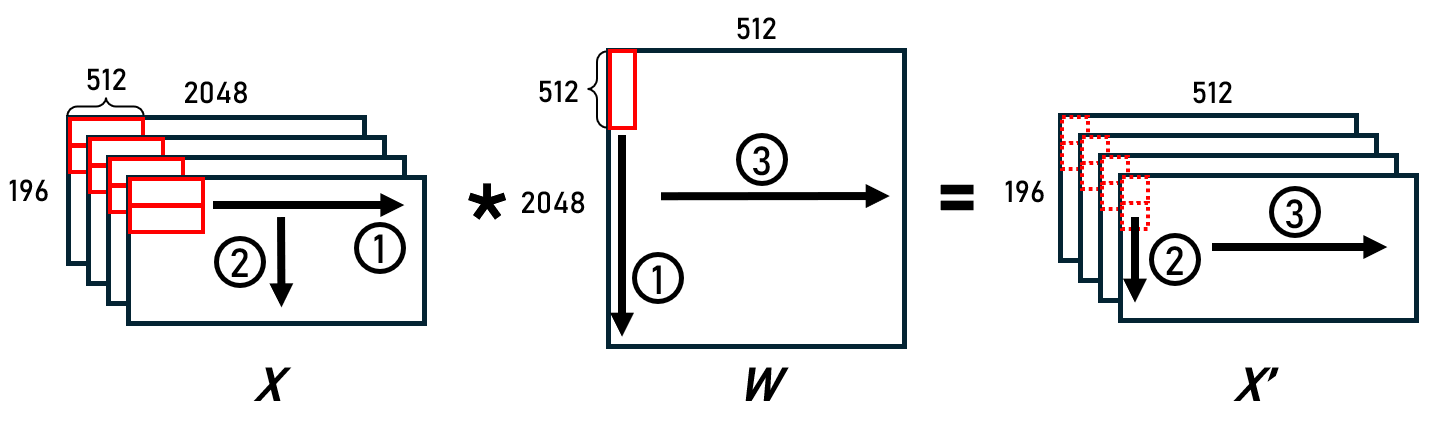}
\caption{Modified Weight Stationary Spiking Linear Operation (WSSL) for MLP2 layer, splitting long columns into segments of 512 to fit into the PE module.}
\label{linear_layer_2048}
\end{figure}

In the "Spikformer V2-8-512-IAND" model, the encoder blocks, which form the backbone of the architecture, each comprise a Spiking Self Attention (SSA) module and an MLP block. While the SSA module involves dot-product computations for producing $QK^{T}V$, the majority of computations within the eight encoder blocks are dedicated to linear layers. These linear layers do not benefit from the inherent weight-sharing characteristic of convolution layers, presenting unique challenges for computational optimization.

As illustrated in Fig.~\ref{linear_layer}, the architecture employs the Weight Stationary Spiking Linear Operation (WSSL) for linear layers with a height of 512. WSSL iterates through the entire input map across four timesteps to produce one complete column in the output map before progressing to the next column of the weight map, continuing this process until the entire output map is complete.

For the MLP2 layer in the encoder block, which has a dimension of 2048x512, the weight matrix column is strategically divided into four sections to fit into the 512 PE units. This segmentation facilitates the computation of outputs for two pixels across four timesteps by sequentially combining the contributions from each of the four cycles. As illustrated in Fig.~\ref{linear_layer_2048}, the modified WSSL ensures that a complete output is formed for each pixel while optimizing memory usage, requiring only an additional 192-bit buffer to store intermediate results from the first three cycles.

\subsection{Spiking Tile-wise Dot Product Calculation (STDP)}
\begin{figure}[t]
\centering
\includegraphics[width=0.7\linewidth]{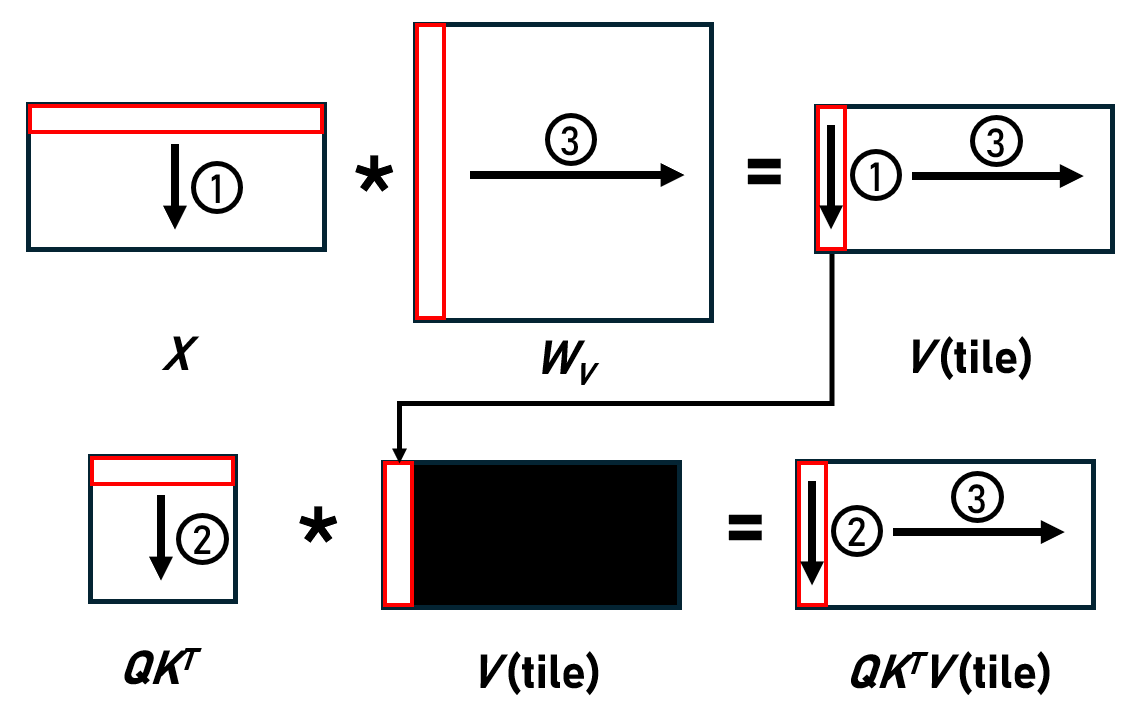}
\caption{Spiking Tile-wise Dot Product Calculation (STDP)}
\label{dot_product}
\end{figure}

Inspired by the tile-wise fused linear-attention-linear context matrix calculation in \cite{ULSeq-TA}, we simplified its structure to better leverage the spike-form inputs of the Spiking Self Attention (SSA) module, leading to the development of the Spiking Tile-wise Dot Product Calculation (STDP). By computing the dot-product of \( QK^T \) with each column of matrix  \( V \) immediately after the column is completed, as illustrated in Fig.~\ref{dot_product}, STDP eliminates the need to store the entire matrix  \( V \). This approach allows us to temporarily hold only one column of  \( V \), which is discarded right after the corresponding column of \( QK^TV \) is computed. Consequently, this method reduces memory usage and enhances computational efficiency.

\section{Experimental Result}

\subsection{Hardware Implementation Result}

The proposed accelerator, VESTA, has been designed using System Verilog and synthesized with TSMC 28nm CMOS technology using Synopsys Design Compiler. VESTA operates at a clock frequency of 500MHz and can execute the "Spikformer V2-8-512-IAND" model at 30 frames per second for 224×224 images with 3 channels (RGB) for image classification. Since VESTA is built upon the Spikformer V2 architecture, which is designed for classifying the ImageNet dataset, VESTA is capable of classifying images into 1000 categories.

The core area of VESTA is \(0.844mm^2\), with a gate count equivalent to 523k gates for the given technology, and it utilizes a total of 107KB of SRAM. To achieve highly parallel computation, the PE module contains 512 PE units, each consisting of 8 PE blocks. Consequently, the PE module occupies 52.92\% of the total core area. The adder tree, which sums the different forms of output from the PE module, accounts for 40.41\% of the core area. The TFLIF module occupies 5.73\%, while other components together take up 0.94\% of the core area. %

Compared to prior SNN hardware accelerators, as summarized in Table~\ref{Comparison}, VESTA demonstrates superior performance with a smaller SRAM size, much higher peak throughput, and notable improvements in both area and energy efficiency.

\begin{table}[!ht]
\caption{Comparison with Prior Work}
\label{Comparison}
\centering
\begin{tabular}{|l|c|c|c|}
\hline
                            & \textbf{This work}   & \cite{TCAS} & \cite{SpinalFlow}      \\ \hline
Technology                    & 28nm        & 28nm  & 28nm  \\ \hline
Voltage   (V)                 & 0.9         & 0.81  & -     \\ \hline
Frequency   (MHz)             & 500         & 500   & 200   \\ \hline
Network   Architecture in SNN & Transformer & CNN   & CNN   \\ \hline
Spatial-Temporal   Processing & Yes         & No    & No    \\ \hline
Weight   Precision (bits)     & 8           & 8     & 8     \\ \hline
PE number                     & 4096        & -     & 128   \\ \hline
SRAM (KB)                     & 107         & 240   & 585   \\ \hline
Peak   Throughput (GSOPS)     & 4096        & 1150  & 51.2  \\ \hline
Core Area   (mm\(^2\))        & 0.844       & 0.89  & 2.09  \\ \hline
Area efficiency   (TSOPS/mm\(^2\))  & \textbf{4.855}       & 1.292 & 0.024 \\ \hline
Core Power   (mW)             & 416.1       & 149.3 & 162.4 \\ \hline
Energy efficiency   (TSOPS/W)        & \textbf{9.844}       & 7.703 & 0.315 \\ \hline
\end{tabular}

\vspace{5pt}
\begin{minipage}{0.89\linewidth}
\raggedright
\footnotesize{GSOPS = Giga spike operations per second}\\
\footnotesize{TSOPS = Tera spike operations per second}\\
\end{minipage}
\end{table}

\subsection{Computation Time Distribution of Each Operation}

Although the Self Attention (SSA) module is the core of a transformer-based model, in the "Spikformer V2" model we are implementing, the MLP module in the Spikformer Encoder Block takes up most of the cycles due to the large size of the linear layers. As a result, the Weight Stationary Spiking Linear Operation (WSSL) occupies 80.79\% of the total computation time. The computation time distribution for the other operations is depicted in Table~\ref{percentage_time}.

\subsection{Benefits of Proposed Methods}

VESTA employs optimized methods for each operation in the Spikformer V2 model. By improving PE module utilization rates and reducing buffer sizes, we can enhance energy efficiency and decrease the total computation cycles required. The benefits of each proposed method are summarized in Table~\ref{benefits}.

\begin{table}[t]
\caption{Summary of Computation Time Distribution \\for Each Proposed Method}
\label{percentage_time}
\centering
\begin{tabular}{|c|c|}
\hline
Proposed Method & Percentage \\ \hline
Zig-Zag Spiking Convolution            & 0.19\% \\
Shift-and-Sum Spiking Convolution    & 4.13\%  \\
Weight Stationary Spiking Linear Operation     & 80.79\% \\
Spiking Tile-wise Dot Product Calculation    &  14.88\%\\
\hline
\end{tabular}
\end{table}

\begin{table}[t]
\caption{Summary of Benefits for Each Proposed Method}
\label{benefits}
\centering
\begin{tabular}{|c|c|c|}
\hline
\multicolumn{1}{|c|}{\begin{tabular}[c]{@{}c@{}} Proposed\\ Method\end{tabular}} & \multicolumn{1}{c|}{\begin{tabular}[c]{@{}c@{}}Improve PE module \\ Utilization Rate\end{tabular}} & \multicolumn{1}{c|}{\begin{tabular}[c]{@{}c@{}}Reduce  Buffer \\ Size\end{tabular}} \\ \hline
ZSC                    & v                                                                                                  & v                                                                                        \\ \hline
SSSC                   & v                                                                                                  &                                                                                          \\ \hline
WSSL                   &                                                                                                   & v                                                                                        \\ \hline
STDP                   &                                                                                                   & v                                                                                        \\ \hline
\end{tabular}
\end{table}

\section{Conclusion}

In this paper, we introduced VESTA, a versatile hardware accelerator designed to integrate Spiking Neural Networks (SNNs) with transformer architectures. VESTA addresses the computational and power challenges of deploying transformers on edge devices by unifying processing elements (PEs) to efficiently handle convolutional, linear, and dot product operations using spike-form data.

Innovations like Zig-Zag Spiking Convolution (ZSC), Shift-and-Sum Spiking Convolution (SSSC), and Spiking Tile-wise Dot Product Calculation (STDP) optimize computational efficiency and memory usage. The Temporal Fused Leaky Integrate-and-Fire (TFLIF) module further enhances the architecture's efficiency by integrating batch normalization processes and reducing memory overhead. By employing these strategies, VESTA achieves a substantial reduction in complexity and energy usage, setting a new benchmark for energy-efficient hardware accelerators suitable for edge inference applications. 

\nocite{zhou2022spikformer}
\nocite{zhou2023spikingformer}
\nocite{dosovitskiy2020image}
\nocite{MAASS19971659}
\nocite{li2022spikeformer}
\nocite{Time_Batching}
\nocite{Sparse_Compressed_SNN}

\bibliographystyle{IEEEtran}

\bibliography{IEEEabrv,bib/ieeeBSTcontrol,bib/thesis}

\end{document}